\documentclass[twocolumn,showpacs,preprintnumbers,amsmath,amssymb]{revtex4} 
\usepackage{graphicx}% Include figure files
\usepackage{dcolumn}% Align table columns on decimal point
\usepackage{bm}% bold math
\usepackage{color}

\def\ni{\noindent}
\def\br{\vec{r}}

\def\div{\vec{\nabla}}

\def\sig{{\sigma}}

\def\ni{\noindent} 
\def\beq{\begin{equation}} 
\def\eeq{\end{equation}}

\begin{document}
\draft

\title{Granular solids transmit stress as two-phase composites}

\author{Raphael Blumenfeld}
\email[]{rbb11@cam.ac.uk;  
https://orcid.org/0000-0001-7201-2164}
%\homepage[]{Your web page}
%\thanks{}
\affiliation{Gonville \& Caius College, University of Cambridge, Trinity St., Cambridge CB2 1TA, UK}

\date{\today}

\begin{abstract}

A basic problem in the science of realistic granular matter is the plethora of heuristic models of the stress field in the absence of a first-principles theory. Such a theory is formulated here, based on the idea that static granular assemblies can be regarded as two-phase composites. 
A thought experiment is described, demonstrating that the state of such materials can be varied continuously from marginal stability, via a two-phase granular assembly, then porous structure, and finally be made perfectly elastic.  For completeness, I review briefly the condition for marginal stability in infinitely large assemblies. 
The general solution for the stress equations in $d=2$ is reviewed in detail and shown to be consistent with the two-phase idea.
A method for identifying the phases of finite regions in larger systems is constructed, providing a stability parameter that quantifies the `proximity' to the marginally stable state.
The difficulty involved in deriving stress fields in such composites is a unique constraint on the boundary between phases and, to highlight it, a simple case of a stack of plates of alternating phase is solved explicitly. An effective medium approximation, which satisfies this constraint, is then developed and analysed in detail. 
This approach forms a basis for the extension of the stress theory to general granular solids that are not marginally stable or at the yield threshold.

\end{abstract}

%\pacs{64.60.Ak, 05.10.c 61.90.+d}

%\keywords{granular solids, marginal rigidity, isostaticity, coordination number, non-convex particles}

\maketitle

\section{Introduction} 

Granular matter (GM), whose ubiquity on Earth is second only to water, is essential not only to human society but also to most life on land. It is often regarded as a distinct form of matter because of its rich behaviour, which is dissimilar from the conventional forms of matter. 
Of essential importance is understanding and predicting how GM transmits stress. A first-principles stress theory in these materials is essential in a wide range of disciplines: civil, structural, and chemical engineering; geology and Earth sciences, and physics, as well as in technological applications of powders, soils, foodstuff, etc. It is also key to mitigation of hazards, from snow and soil avalanches to deflecting rubble-pile asteroids.  

The science of GM is at least two thousands and two hundred years old. Indeed, what is regarded today as the oldest existing scientific publication, dating back to the third century BCE~\cite{Archimedes}, involved GM.To an extent, this is attestation of the significance of this field. In the late 19th century~\cite{Re1885} and in the early 20th century~\cite{Ba41}, work on GM was motivated by practical applications and was mainly done within the context of engineering.The last three decades saw an explosion of fundamental theoretical research, following the seminal work of Edwards~\cite{EdOa89a,EdOa89b,MeEd89}. 
Yet, in spite of this uniquely long history and intensified recent research activity, no first-principles stress theory for such media exists. 
One of the reasons is that, unlike any conventional continuum, GM behaves as a combination of a solid and a fluid and it transmits stress very non-uniformly, often via stress chains~\cite{SeWu46,Wa57,Da57,BaKi85,HoBe97,OdKa98,Vaetal99,Maetal07}. Another reason is that there is a range of phenomenological and empirical models, utilised in engineering, providing the impression that one can get away without a fundamental theory. This situation is unsatisfactory and, indeed, subsidence and collapses of buildings and structures provide evidence that, while useful, empirical models have serious limitations. 

It has been suggested that one of the hurdles to constructing such a stress theory is that GM is regarded paradigmatically as a continuum endowed with some constitutive properties, for which stress equations need to be developed. Since this approach has not been fruitful for many decades, it was proposed that general GM need to be regarded rather as two-phase composites, with each phase satisfying different stress field equations~\cite{Bl04}. It is this view that I intend to explore in the following.

Specifically, several arguments are presented in support of the two-phase-composite idea and a simple case of such a composite is solved. A method to derive the stress from first-principle in such media, using an effective medium approach, is formulated. To alleviate a difficulty in distinguishing between the different phases visually, which is important for the purpose of imposing boundary conditions on the phase boundaries, a quantitative stability parameter is developed, which can also be used as a phase field parameter. To make this paper self-contained, I also review briefly: (i) the method of identifying marginally stable granular assemblies and (ii) the current isostaticity stress theory (IST) for the marginally stable state of GM, with a specific solution in two dimensions ($d=2$). 

The structure of the paper is the following.
In section {\bf II}, the state of marginal stability of GM is defined quantitatively in terms of the particle-scale mean coordination number (MCN). 
In section {\bf III},  the existing stress theory for marginally stable GM is reviewed briefly, with more details, including the general solution in two-dimensional systems, given in the supplemental material\cite{SM1}. 
In section {\bf IV}, I discuss the role of the marginally stable state as a critical point in the traditional sense, with a proper diverging response length, which is reflected in the increasing typical length of force chains. This state, which is also the yield threshold, is often referred to as a critical state in the engineering literature, albeit without the connotation that this term usually carries in physics.
A thought experiment is then described, which illustrates clearly that GM is a two-phase composite, with one phase isostatic and the other elastic. The larger the concentration of the former phase the longer the response length. 
In section {\bf V}, the construction of a general stress theory for such two-phase composites is discussed. 
An example of a simple case, in which alternate-phase plates are arranged in series is analysed, solved exactly, and used to illustrate a fundamental difficulty, which can be traced back to the assumptions of isostaticity theory. 
Then a possible extension by an effective medium method is described and the difficulties posed by a more general theory are discussed. 
In section {\bf VI}, a stability parameter is introduced, which makes possible a local quantitative distinction between the phases in finite granular regions. This parameter also enables a quantitative determination of the `distance' from the critical point. 
Finally, the results and some implications are discussed in the concluding section {\bf VII}. 

%===================================

\section{The marginally stable state} 

At the macroscopic, many-particle level, the marginally stable state is the (macro-)state at the yield threshold between the fluid and solid states. It is also known as critical, marginally rigid, and isostatic state. The reason that this is the yield threshold can be traced to the particle level, at which the number of force-carrying inter-particle contacts is such that the number of equations to determine the inter-particle forces is exactly equal to the number of unknown force components that require determination. When there are too few such contacts, the medium is unstable and must rearrange under external forces. 
This state is marginally stable because any perturbation in the applied load or a particle's position, gives rise to contact breaking and to local rearrangement. This perturbs neighbour particles and so on. Thus, a perturbation of one contact can lead to a rearrangement of a significant portion of the granular assembly. Such a long range response to a perturbation is the hallmark of a critical point, as will be discussed below.

The difference between the numbers of unknowns and balance equations to determine them, is quantified by the mean coordination number (MCN), $z$, which is defined as the number of force-carrying contacts per particle. The marginally stable  state corresponds to a `critical' value, $z_c$, which depends on: the dimensionality, $d$; whether the particles are frictional or are frictionless; and whether they are perfectly circular/spherical/hyper-spherical or of other shapes. When $z<z_c$, the medium is fluid and when $z>z_c$ it is solid. 

To determine $z_c$, we need to consider $d$-dimensional many-particle assemblies of $N\ (\gg1)$ rigid particles of convex shapes. It is straightforward to extend the discussion to some classes of non-convex shapes and to compliant hard particles, but this would add very little insight and this issue is better circumvented here. 
In the following analysis, only fixed compressive boundary forces are presumed to act on the granular assemblies -- external force fields, including gravity, are ignored. The justification for this is that given a static structure of an assembly, the stress equations discussed below are linear, which means that the effects of an external force field can be superposed on the IST solution. \\

\noindent{\bf Frictional particles}:\\
Frictional particles experience $d$ force components at each contact point, which need to determined. Neglecting boundary effects for very large assemblies, summing over the coordination numbers around all particles results in twice the total number of contacts, $C_d$, namely, $C_d=Nz/2$. There are therefore $dNz/2$ unknowns. To be mechanically stable, each particle must satisfy $d$ conditions of force balance and one torque balance condition for each of the $d(d-1)/2$ axes of rotation. The critical MCN must then satisfy the equality
\beq
d\frac{z_c}{2} N = \left[ d + \frac{d(d-1)}{2}\right] N \quad \Rightarrow \quad z_c = d + 1 \ .
\label{ZcFriction}
\eeq
This calculation can be found extensively in the literature.\\

\noindent{\bf Frictionless non-(hyper-\!)spherical particles}:\\
In this case, the force must be normal to the tangent plane at the contact point and, therefore, the geometry determines the direction of any contact force. This leaves only one unknown per contact -- the force magnitude. The number of unknowns is then, $C_d=z_cN/2$. The number of equations is the same as on the right hand side of eq. (\ref{ZcFriction}) and equating it with the number of unknowns yields 
\beq
z_c = d(d + 1) \ .
\eeq

\noindent{\bf Frictionless hyper-spherical particles}:\\
An assembly of frictionless perfect hyper-spheres, which includes discs in $d=2$, is often used in numerical simulations because it is convenient for contact detection and contact force transmission. However, not only is it difficult to reproduce physically, but such an assembly is also degenerate in the sense that balance of forces on every particle ensures automatically balance of torques. Therefore, the torque balance conditions are redundant for all particles and only the $Nd$ force balance conditions must be satisfied. Since, for  such particles, the forces are also normal to the contact tangent plane, there is only one unknown to determine at each of the $z_cN/2$ contact points. Equating unknowns and equations then yields
\beq
z_c = 2d \ .
\eeq

It should be commented that the values for $z_c$, calculated for all types of particles, incur boundary corrections of order ${\cal{O}}\left(N^{-1/d}\right)$, which have been neglected. These corrections will become relevant for the discussion in section {\bf VI}.

%===================================

\section{Critical stress transmission at marginal stability}

As mentioned, force chains are the conduits of stress and displacement perturbations and the longer they are the further the response. In particular, in the marginally stable state the typical length of force chains is comparable to the system size, making this state the equivalent of a conventional critical point. This equivalence is key to understanding stress transmission in more general states of GM. It is therefore useful to review briefly the theory of stress transmission at marginal stability.  

Any continuum stress theory must satisfy the balance conditions:
\begin{align}
\div\cdot\overline{\overline{\sigma}} &= \vec{g}_{ext} \quad\quad ({\rm balance\ of\ forces})\\
\overline{\overline{\sigma}} &= \overline{\overline{\sigma}}^T \quad\quad ({\rm balance\ of\ torques})\ .
\label{Balance}
\end{align}
In $d$ dimensions, the first equation provides $d$ conditions, the second $d(d - 1)/2$, and together $d(d + 1)/2$ conditions in total. Since the stress tensor has $d^2$ components, further $d(d-1)/2$ equations are required to determine it. These 'closure' equations need to be provided by constitutive relations. In elasticity theory, the closure is by St. Venant's compatibility constraints on the strain tensor, augmented with stress-strain relations~\cite{Mu63}. 
Such closure, however, is not appropriate for the marginally stable state. This is because the stress field is nothing but a continuum representation of the spatial distribution of inter-particle forces in the marginally stable state and, since these forces are exactly determinable by the structure and are independent of any infinitesimal displacement that led to it, then the continuum stress cannot depend on the strain field. This is also evident from the fact that no elastic moduli are involved in the above discussion of the determination of those forces.  
It follows that the only relevant constitutive characteristics must be based on the local structure. The observations of non-uniform stress transmission in GM via chains~\cite{SeWu46,Wa57,Da57,BaKi85,HoBe97,OdKa98,Vaetal99,Maetal07} further supports the idea that the equations cannot be elliptic and therefore cannot arise from strain-based constitutive relations. 
It was proposed then that the closure is by a stress-structure relation~\cite{Wietal96,Wietal97,Caetal98a,Caetal98b},
\beq
\overline{\overline{M}}:\overline{\overline{\sigma}} = 0 \ ,
\label{StressStructure}
\eeq
in which $\overline{\overline{M}}$ is a symmetric tensor that characterises the local structure. Its determinant is negative, which results in {\it hyperbolic} equations, in contrast to the elliptic equations of elasticity theory. This gives rise to solutions that `propagate' into the medium along characteristic paths. Along these paths, which can be interpreted as stress chains, characteristic stress combinations are constant. The set of equations (\ref{Balance}) and (\ref{StressStructure}) are commonly called {\it isostaticity theory}. So far, the tensor $\overline{\overline{M}}$ has been derived from first-principles only in $d=2$~\cite{BaBl02,Bl04,Geetal08,BlMa17}. Nevertheless, there is a range of empirical models for it, or leading to it, in $d=2$ and $3$, e.g., Mohr-Coulomb~\cite{JaCo79}, Tresca~\cite{Tr64}, von Mises~\cite{vM13}, Drucker–Prager~\cite{DrPr52}. The characteristics can be straight, curved and even bend backwards~\cite{BlMa17}. A brief outline of the solution of these equations in rectangular coordinates and an example of a solution are given in the supplemental material\cite{SM1}. 
It should be commented, that the first-principles theory holds for compliant particles, as long as the MCN is $z_c$ and the compressed areas at contacts are small compared to the particle sizes. Compliance introduces corrections to the solutions of eqs. (\ref{Balance})-(\ref{StressStructure}), which decay as the number of particles increases~\cite{NonRigid}. 

The marginally stable state acts as a critical point in that a small displacement of a particle can lead to the yield of large part of the assembly~\cite{Maetal07,Reetal13,Duetal14,Deetal14}. The main descriptor of this state is the critical MCN, $z_c$, and the deviation from this state can be parameterised by the difference $z - z_c$. The critical nature of the marginally stable state opens the door to modelling GM in general, which is the subject of the next section. \\
It should be commented in passing that, while it is tempting to consider the typical length of the characteristics stress chains as a descriptor of the long-range correlation, this similarity holds only for uniform fabric tensors, $\overline{\overline{M}}$. This is because stress chains straight in such systems and span the entire system. However, when $\overline{\overline{M}}$ is non-uniform, the stress along the characteristics decays with distance and so do the effects of local perturbations. 
There is some fundamental difference between the force chains, observed in experiments with photoelastic particles~\cite{HoBe97,MaBe05,Zhetal08,Zhetal10,Daetal17,Zhetal19}, and stress chains. The former are observed only when above some threshold and, therefore, the definition of a force chain is not sharp to some extent. In contrast, theoretical stress chains are defined uniquely and unambiguously, given the fabric tensor $\overline{\overline{M}}$.

%===================================

\section{General GM is a two-phase composite}

While isostaticity is an established first-principles theory, marginally stable states are rare in realistic static systems, requiring specialised dynamics to generate them. The MCNs of most solid granular assemblies, whether natural or man-made, often exceed $z_c$. The question is how to extend the isostaticity stress theory to such media. To this end, it has been proposed that, at least sufficiently close to the marginally stable state, realistic GM must be regarded as composites comprising regions of two phases: one marginally stable and the other is over-connected, in which $z>z_c$~\cite{Bl04}. The usefulness of the {\it two-phase composites} picture can be illustrated with the following thought experiment.

Consider a large assembly of elastic particles, e.g., rubber balls, initially at a marginally stable state under some infinitesimally small boundary forces. Under such loading, the contact areas can be made much smaller than the smallest ball diameter and isostaticity theory provides the correct solution for the stress field. Now, increase all the boundary forces uniformly by a factor $\alpha=1+\epsilon$, with $0<\epsilon$. When $\epsilon$ is sufficiently small, such that it cannot bring even the closest pair of particles into contact, the number of contacts remains the same and only their areas increase as they are compressed slightly. In a very large assembly, this has been shown only to introduces small corrections to the original solution, with the corrections decaying with system size. As $\epsilon$ increases, new contacts are made here and there and the MCN starts to increases: $z = z_c + \delta z$. When $\delta z\ll1$, the  over-connected regions are small and isolated. A force chain incident on such a region 'scatters' in the sense that its continuation is shared by more contacts than required for marginal stability. This sharing means that each of the forces emerging from this region is lower in magnitude. Setting the magnitude observation threshold of force chains appropriately, the incident force chain effectively 'terminates'. As $\alpha$ increases, more over-connected regions form, the typical length of force chains decreases, and with it the stress. 
This resembles strongly the behaviour of traditional systems as they move away gradually from critical points. For example, increasing the temperature slightly above the critical point introduces regions of normal conductivity, or increasing the concentration of non-conducting elements at the percolation threshold through an otherwise conducting system, reduces the conductivity by generating non-conducting regions. 

Another effect of increasing $\alpha$ is that contact areas between particles in contact increases. When the size of such a contact becomes comparable to the size of either of the particles sharing it, this pair can no longer be regarded as two separate particles. As balls get squeezed together and the contact areas of sufficiently many reach this limit, the assembly can no longer be regarded as granular and is, rather, a porous medium, comprising an elastic solid phase and cavities, or pores. 
Some models for computing stress transmission in this type of media exist~\cite{MoWa02,Laetal17}, but discussing them is tangential to this presentation. 
Finally, at some large value of $\alpha$, these voids are also squeezed out completely and the system becomes a continuous uniform elastic solid. 
The stress fields in such a solids are readily calculated by conventional elasticity theory.

This thought experiment shows that there is a continuous spectrum of structures with the marginally stable critical point at one end and a perfectly elastic state at the other. General GM is on this spectrum sufficiently close to the former, before the appearance of porous media. 
In particular, where on this spectrum a granular solid exactly is depends on the difference $\delta z = z - z_c$, which is tantamount to saying that it depends on the response length.
 
It is clear that, in assemblies of particles that are not as elastic as rubber balls, other physical mechanisms may intervene before the porous medium state or the continuum are reached, such as particle fragmentation, phase transitions, etc. These are all ignored because they are irrelevant to the purpose of this thought experiment. Additionally, if the original particulates are made of non-elastic materials, the stress transmission in the final continuous phase need not satisfy the equations of elasticity theory. All these side issues notwithstanding, starting from a perfectly elastic final state is a useful first step toward a more general theory. The two-phase idea may also provide insight into the observation of two distinct sets of force chain networks in simulations of GM~\cite{Raetal98}.
In any case, this conceptual picture suggests a strategy to extend the theory beyond the ideal marginally stable limit and this strategy is discussed next. 

%===================================

\section{Toward a continuum stress theory of general GM}

Field theories of two-phase composites are generally difficult to construct except when the phases have a special spatial distribution. The main existing methods for arbitrary spatial distributions are effective medium approximation, mean field theory, and renormalisation near critical points. Each of these methods involves some special assumptions. Unfortunately, none of these models can be applied directly to GM composites because they are based on the assumption that the two phases satisfy the same field equations and they differ only by their constitutive properties. Example are: mixtures of two conducting materials, in which both phases obey Ohm's law, but have different conductivities; composites of elastic materials, which are often presumed to obey the same stress equations, but with different elastic moduli; and mixtures of dielectrics having electric-displacement fields relation of the same functional form, but with different dielectric constants. 
The two-phase GM problem is more difficult because the phases differ not by their constitutive properties but by the stress equations that they satisfy. This problem is exacerbated by the fact that the elastic phase satisfies elliptic equations and the marginally stable phase, satisfies hyperbolic equations. While the former can be solved under Dirichlet boundary conditions, the latter can be ill-posed under such conditions. Thus, much care is required even in posing the problem.  \\

\begin{figure}[h]
\includegraphics[width=8cm]{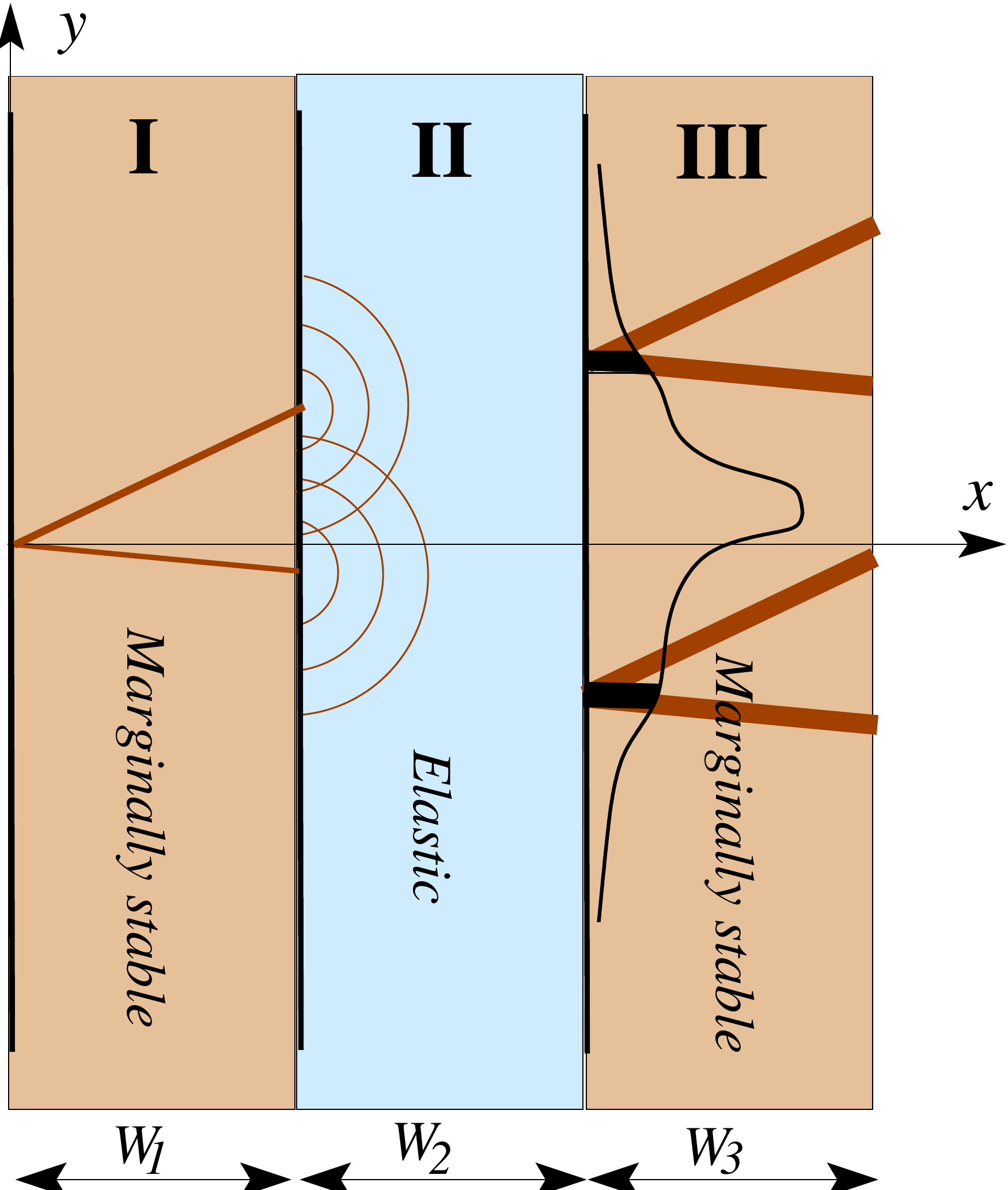}
\caption{A stack of alternating marginally stable and elastic plates. A localised stress is applied at the boundary $x=0$, generating two stress chains that 'propagate' along two characteristic paths. The chain stresses apply two localised loads on on the strain-free boundary at $x=W_1$. The boundary at $x=W_1$ deforms to transmit this stress to the elastic plate. The stress response within the elastic plate satisfies the elliptic equations of elasticity theory. The stress response on the strain-free boundary at $x=W_2$ is sketched. Adding another plate of marginally stable medium at $x= W_1+ W_2$, the stress solution within it is a superposition of the stress chains, which emanate from every point along this boundary, such as the two exemplified in the figure.  }
\label{IsoElastLayer}
\end{figure}

\noindent{\bf Isostatic-elastic pair of plates}:\\
To illustrate the complexity of the problem, it is useful to start with a simple solvable structure in two dimensions. 
Consider only the two parallel plates, {\cal{I}} and {\cal{II}}, sketched in Fig.~\ref{IsoElastLayer}. 
Plate {\cal{I}} is isostatic, occupying $0<x<W_1$ and $-\infty<y<\infty$, and plate {\cal{II}} is elastic, occupying $W_1<x<W_2$  and $-\infty<y<\infty$. The boundary at $x=W_2$, which also extends to $\pm\infty$ in the $y$-direction, is rigid and stress is not transmitted between plates {\cal{II}} and {\cal{III}}.\\

The equations of both elasticity and isostaticity are linear, given the respective constitutive properties, and it is sufficient to consider a point loading applied to the leftmost plate at the origin, $\overline{\overline{\sigma}}(x=0,y=0)$. A more general loading is the superposition of such point loadings. The full solution to the point loading problem is detailed in the supplemental material\cite{SM1}. To summarise it, the stress field response in the marginally stable region {\cal{I}}, whose example structure tensor is chosen to be uniform, for simplicity, $\overline{\overline{M}} =
\begin{pmatrix}
3 & 1\\
1 & -1
\end{pmatrix}$, consists of a finite stress only along two straight stress chains. The gradients of the stress chains are $\lambda_1=3$ and $\lambda_2=-1$ and they follow the {\it characteristic} paths. A long each path, the stress field is a characteristic combination of the stress components that originate from the source at $(x=0,y=0)$. Outside these paths, the stress is exactly zero. This solution superposed with the uniform stress field due to the uniform loading on the boundary, which is also detailed in the supplemental material\cite{SM1},  
\begin{align}
\overline{\overline{\sig}}_{uniform} = 
\begin{pmatrix}
\sig_{xx} & \sig_{xy} \\
\sig_{xy} & \sig_{yy}=3\sig_{xx}+2\sig_{xy}
\end{pmatrix} \ .
\label{UniformStress}
\end{align}
The value of the loading $\sig_{yy}$ must depend on the values of $\sig_{xx}$ and $\sig_{xy}$ to satisfy the constitutive stress-structure relation (\ref{StressStructure}).\\

The stress  chains of the solution are incident on the boundary between regions {\cal{I}} and {\cal{II}}, $x=W_1$, giving rise to two point loadings on this boundary at $y=-W_1$ and $y=3W_1$, 
\begin{align}
\overline{\overline{\sigma}}_1\left(W_1,3 W_1\right) &=  \frac{\sig_{xx} + \sig_{xy}}{4}
\begin{pmatrix}
1 & 3 \\
3 & 9
\end{pmatrix} \notag \\
\overline{\overline{\sigma}}_2\left(W_1,-W_1\right) &= \frac{3\sig_{xx} - \sig_{xy}}{4} 
\begin{pmatrix}
1 & -1 \\
-1 & 1
\end{pmatrix} \ .
\label{SigW1}
\end{align}

The boundary condition at $x=W_1$ must be considered carefully now. If this boundary is presumed to remain straight and independent of $y$, then the stresses at the points $y=-W_1, 3W_1$ along this boundary, would not be transmitted to the elastic medium. Some boundary deformation is required for that. The problem is that isostaticity theory does not provide a way to predict this deformation because strain plays no role in it.
Nevertheless, such a deformation will occur because the application of the load at $(0,0)$ changes the structure wherever the stress is finite. This issue and its effect on the choice of this boundary condition are discussed in some detail in the concluding section, a discussion that touches on the assumptions underlying isostaticity theory. To summarise it here, since there is currently no theory to predict local structural changes as a function of the local stress perturbation, the only way forward is to impose a boundary condition at $x=W_1$ that transmits faithfully the stress from left to right. The natural way to achieve that is to impose a deformation, or strain, $\overline{\overline{e}}$, that satisfies the stress-strain relation in the elastic medium, namely, 
$\overline{\overline{\sigma}}(x=W_1^-,y)=\breve{C}_{II}\overline{\overline{e}}(x=W_1^+,y)$, with $\breve{C}_{II}$ the fourth-order stiffness tensor of the elastic medium in {\cal{II}}.
Applying this boundary condition to the problem at hand, the stress at the left boundary of plate {\cal{II}} comprises two $\delta$-functions, as sketched in Fig.~\ref{IsoElastLayer}, and, together with the condition of a flat rigid boundary on the right of region {\cal{II}}, make for a well-define formulation for the solution in the elastic plate.  \\

Since the strain at, and therefore the distortion to, the left boundary is known, a convenient way to solve for the stress in this region is to first mapping conformally the physical domain with the distorted boundary to a rectangle. Solve for the stress in the mapped domain, using textbook methods~\cite{Muskh}, and then transform the solution back to the physical plane. Two such point-loading solutions are sketched in the figure. \\

For completeness, it should be commented that, when the fabric tensor $\overline{\overline{M}}$ is not uniform in the marginally stable plate, secondary paths of lower stresses emanate from the main characteristic paths, which reach the boundary at $x=W_1$ at different locations. These modify the boundary stress for the elastic plate in a manner that can also be calculated from the solution in the supplemental material\cite{SM1} and can be treated as a superposition of source points at $x=W_1$. \\

{\bf A chain of alternating-phase plates}:\\

Next, consider a longer chain of parallel plate of alternating phases, by adding them to the right of plates {\cal{I}} and {\cal{II}}. The first of this chain, {\cal{III}}, is shown in Fig.~\ref{IsoElastLayer}. They have different thicknesses and all similarly extend to $\pm\infty$ in the $y$-direction. 
Applying the same source load at $(x=0,y=0)$, the stress response in plate {\cal{I}}, as well as its transmission across the boundary at $x=W_1$, are the same as for the pair system discussed above. The boundary condition at $x=W_2$ is straightforward to determine: since the marginally stable medium in plate {\cal{III}} is rigid, it is chosen to be flat. Then, the solution in {\cal{II}} is the same as in the pair system and, consequently, so is the stress at 
$\overline{\overline{\sigma}}(x=W_2^-,y)$. This boundary stress is transmitted to the medium in {\cal{III}} at $\overline{\overline{\sigma}}(x=W_2^+,y)$. 
Assuming that the fabric tensor in {\cal{III}} is the same as in {\cal{I}}, the conceptual `propagation' from two arbitrary source points along the boundary at $x=W_2$ is exemplified in Fig.~\ref{IsoElastLayer}. Each such point plays the same role as the point load at $(x=0,y=0)$. \\

A consistent set of boundary conditions for a chain of $2N$ such plates is then the following. The boundaries at $x=W_{2k}$($k=1,2,...,N/2$), which transmit stress from the $2k$th elastic plate to the $(2k+1)$th marginally stable one, are presumed rigid and flat, while the boundaries at  $x=W_{2k-1}$, which transmit stress from $(2k-1)$th marginally stable plate to the $2k$th elastic one, deform such that the strain generated by the deformation matches the stress-strain relations in the elastic part, $\overline{\overline{\sigma}}(x=W_1^-,y)=\breve{C}_{2k}\overline{\overline{e}}(x=W_1^+,y)$. \\

\begin{figure}[h]
\includegraphics[width=8cm]{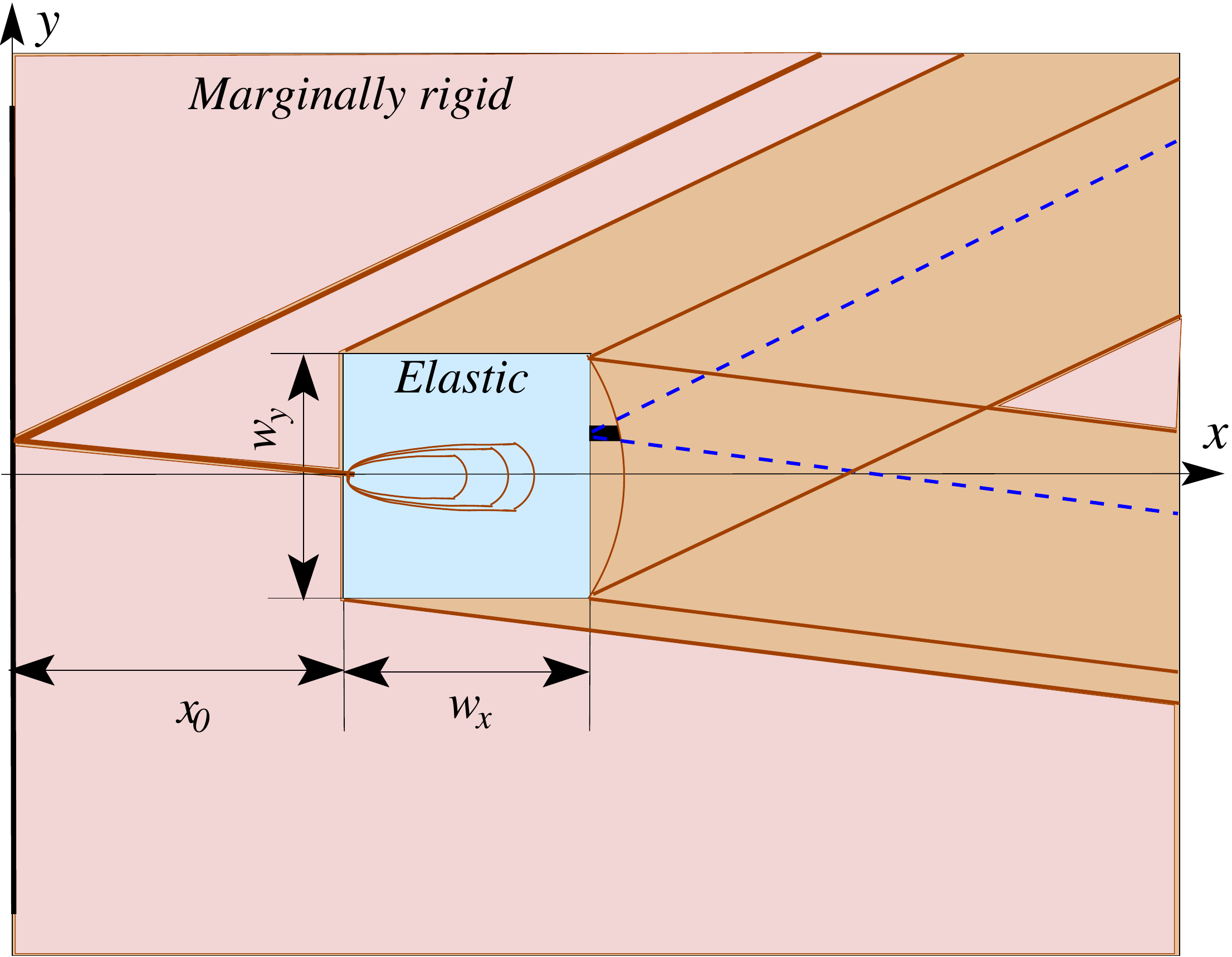}
\caption{A rectangular inclusion (light blue) in an otherwise marginally stable medium (light brown). Stress chains (dark brown) emanate from the point loading at (0,0) along two narrow characteristic path. The chain incident on the elastic inclusion deforms the boundary slightly, 'letting' the stress through and giving rise to an intra-inclusion stress field that satisfies the linear elasticity equations. The inclusion's other boundaries are rigid. The inclusion 'diffracts' the stress, which re-emerges into the marginally stable medium at a much attenuated magnitude along wider paths (dark brown regions).  }
\label{ElasInclusion}
\end{figure}

\noindent{\bf Effective medium method - \\
possibilities and difficulties}:\\

The aim of this subsection is to outline an Effective Medium approximation (EMA) approach for deriving the stress in a general GM composite, rather then develop it in full detail. EMAs are based on the assumption that one phase is sufficiently dilute, often as inclusions, within the other. In this approximation, one neglects the effect of the inclusions on one another. Consequently, the key ingredient in an EMA is then the solution for an isolated inclusion of one phase within an otherwise much larger medium composed of the other phase. By interchanging the roles of the phases, this approach can be applied close to either the marginally stable state or the purely elastic state. 
Analysis of a marginally stable inclusion in an elastic medium is straightforward: the marginally stable medium can be regarded as a rigid inclusion in a large elastic medium, for which solutions exist or can be found with standard elasticity theory~\cite{MaLi21}. \\

The opposite limit, of an elastic inclusion in a marginally stable medium, requires a careful consideration. While diffraction of hyperbolic characteristics from scatterers has been discussed in the literature~~\cite{Cletal98b}, this is less relevant in this context than the stress developing within a finite inclusion. Let the medium occupy the half-space $x>0$ and $-\infty<y<\infty$ and the stiffness tensor within the inclusion be $\breve{C}_{inc}$. For clarity, assume again that its fabric tensor is spatially uniform; as mentioned, position-dependent fabric tensors, $\div\cdot\overline{\overline{M}}\neq0$ lead to non-straight chains, stress attenuation along them, and branching, all of which, although making the treatment more involved quantitatively, can be included without any conceptual difficulty in the following approach.  
It is convenient to consider a rectangular elastic inclusion, as shown in Fig.~\ref{ElasInclusion}. \\

Consider a set of discrete point loadings on the boundary at $x=0$, at intervals $\chi_i$, with $\chi_i$ narrowly distributed around a mean value $\chi_0$. These act as sources and from each one can trace two characteristic paths into the marginally stable medium. The paths from one such source are shown in Fig.~\ref{ElasInclusion}. The characteristic stress component combination on each path is determined by the solution described in the supplemental material\cite{SM1}. In the absence of the inclusion, the stress field inside the medium, $\Sigma_0(x,y)$, consists of a network of stress chains. 
This solution would be unaffected when no chain is incident on the inclusion and the probability for this to happen, $p_0$, decreases with $W_y/\chi_0$, most likely as $e^{-W_y/\chi_0}$ although its exact functional form is immaterial for the present discussion. \\

When a stress chain is incident on the inclusion, which is the case illustrated in Fig.~\ref{ElasInclusion}, it provides a point loading on the boundary of the elastic inclusion at $x=x_0^-$. As illustrated in alternating plates system, the way to transmit the stress to within the inclusion is by posing that this boundary is deformed into the inclusion such that the strain at $x=x_0^+$ satisfies 
$\overline{\overline{\sigma}}(x=x_0^-,y=0)=\breve{C}_{inc}\overline{\overline{e}}(x=x_0^+,y=0)=\overline{\overline{\sigma}}(x=x_0^+,y=0)$. 
Following the example of the system of alternating plates, the boundaries of the inclusion, on which no stress chain is incident, should be regraded as flat and rigid. 
Given these conditions, the stress field inside the inclusion, can be calculated either analytically or numerically, using linear-elasticity. Again, if the calculation with the deformed boundary is problematix, one can conformally map the inclusion back to the original rectangle, solve for the intra-inclusion stress in the mapped plane and then conformally-map this solution back to the physical plane. A schematic illustration of contours of equal-$\overline{\overline{\sig}}_{xx}$ within the inclusion is also shown in Fig.~\ref{ElasInclusion}. 
This calculation then yields the stress distribution along the rigid boundaries, which are then transmitted to the rest of the marginally stable medium. This transmission must follow also the characteristic paths, as sketched in the figure. The `re-emerging' stress paths are broad, corresponding to the size of the inclusion and orientation differences between the boundaries and the two characteristics.  \\

As a consequence of force balance, the stress components magnitudes within the widened stress paths are suppressed to well below those of the original incident chain. Setting a detectability threshold, as for force chains, the stress is likely to drop below the threshold and, to all practical purposes, the incident stress chain effectively terminates at the inclusion. The larger the inclusion the wider the re-emerging paths and the stronger the suppression. Denoting the single-inclusion stress field $\Sigma_1$, the EMA stress field is 
\beq
\Sigma_{EMA} = p_0\Sigma_0 + \left(1 - p_0\right)\Sigma_1 \ .
\label{EMASol1}
\eeq
Placing a second inclusion elsewhere, gives rise to a similar solution, $\Sigma_2$. Since the inclusions are too far to interact, the EMA stress field due to $n$ such inclusions is
\beq
\Sigma_{EMA} = p_0^n\Sigma_0 + \left(1 - p_0^n\right)\sum_{j=1}^n\Sigma_j\left(\br - \br_j\right) \ ,
\label{EMASoln}
\eeq
in which $\br_j$ denotes the position of the $j$th inclusion.
Increasing the concentration of inclusions and/or their sizes, but without violating the effective medium assumption, increases the MCN, $z_c \to z = z_c + \delta z$. An increase in the inclusion concentration also increases the probability of incidence of stress chains on them and effectively terminating. The consequent shortening of the typical length of stress chains with increase of the MCN is indeed consistent with experimental observations~\cite{HoBe99,BeCh19}. This also makes the EMA consistent with the idea that the value of $\delta z$ controls the response length near the marginally stable critical point. Using then $\delta z$ as a measure of the proximity to the critical point, it is tempting to conjecture that the relation between the stress chains typical length, ${\mathcal{L}}_{\sig}$, and the 'distance' from the critical point follows the conventional power-law form: 
\beq
{\cal{L}}_{\sig} \sim \delta z^{-\nu} \quad ;\quad \nu>0 \ .
\label{ChainLength}
\eeq
This form is consistent with experimental observations near the marginal stability point~\cite{Waetal19}, but it depends on more than the typical length of stress chains. This is because nonuniform fabric tensors, in which $\div m_{ij}\neq0$, give rise to coupled characteristics $\omega_i$, which may lead to chains dropping below the threshold and terminating even if without incidence on inclusions~\cite{Geetal08,BlMa17}. These effects are not taken into consideration in (\ref{ChainLength}) and to include them requires quantifying the dependence of this relation on the gradients of the fabric tensor $\overline{\overline{M}}$. Strong gradients could not only lower the pre-factor in (\ref{ChainLength}) but also increase $\nu$, with each of these effects suppressing ${\cal{L}}_{\sig}$ for a given $\delta z$. A full discussion of the effects of structure tensor inhomogeneity is beyond the scope of this work, but it offers an interesting line of future investigation. \\

%===================================

\section{Identifying the phases in the two-phase composites}

To implement the two-phase-composite idea, it is important to have a clear way to identify the boundaries between the phases. This is particularly important in view of the required careful treatment of the boundary conditions. 
Unlike in many traditional two-phase composites, such an identification is not straightforward because the phases are visually very similar. 
The only structural difference between the phases is their connectivities per particle, or specific connectivities. The specific connectivity of a region $\Gamma$ is defined as $\delta z_\Gamma = z_\Gamma - z_{c,\Gamma}$, with $z_{c,\Gamma}$ the critical value of the MCN that makes the region $\Gamma$ marginally stable and $z_\Gamma$ the actual MCN of the particles within $\Gamma$. This value is different from that of the infinitely large assembly, calculated in section {\bf II}, due to the boundary corrections, which are no longer negligible.  \\

\begin{figure}[h]
\includegraphics[width=8cm]{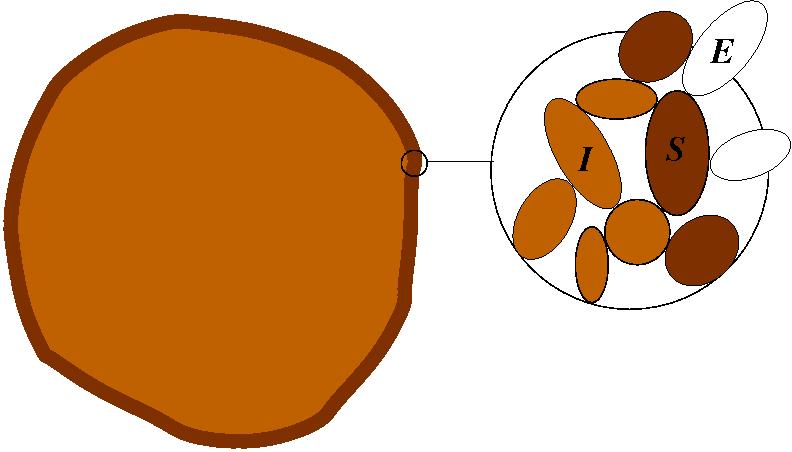}
\caption{
A finite domain $\Gamma$ within a larger granular assembly. The internal particles (light brown, particles labeled 'I') are surrounded by a surface (dark brown, particles labeled 'S'), regarded as its boundary, $\partial_\Gamma$, whose particles are in contact with external particles (white, particles labeled 'E').} 
\label{figGamma}
\end{figure}

A sketch of a finite domain, $\Gamma$, is shown in~Fig.\ref{figGamma}. It contains ${\cal{N}}_\Gamma$ particles, of which ${\cal{N}}_S$ are regarded as its surface and the boundary, $\partial\Gamma$ (dark brown in the figure), between $\Gamma$ and the rest of the assembly. 
Let us define a stability parameter as the difference between the number of unknown force components to determine and balance conditions, per particle in $\Gamma$, 
\beq
J_\Gamma \equiv \frac{\left(N_{unknowns}\right)_\Gamma - \left(N_{conditions}\right)_\Gamma}{{\cal{N}}_\Gamma} \ .
\label{J}
\eeq
Dropping the subscript $\Gamma$, for brevity, the region is unstable and fluid when $J<0$, marginally stable when $J=0$, and stable and solid when $J>0$. The specific connectivity and the stability parameter are equivalent for determining the phase because the number of unknowns is proportional to the number of contacts. The calculation of the stability parameter of $\Gamma$ is done as follows. \\

Within $\Gamma$, there are: $C_{II}$ contacts between internal particles; $C_{IS}$ contacts between internal and surface particles, and $C_{SE}$ contacts between surface and external particles. The external particles exert forces on $\Gamma$ through $\alpha {\cal{N}}_S$ contacts with the surface particles, with $\alpha={\cal{O}}(1)$. The premise is that all these quantities can be extracted visually from $\Gamma$. In the following, I focus on two-dimensional systems, for simplicity, but the analysis can be readily extended to three dimensions. The stability threshold depends on the particle surface friction and whether they are spheres or not. These are discussed next case by case. \\
 
\noindent{\bf Frictional particles in $d=2$}:\\
In the calculation of the MCN of $\Gamma$, the contacts of the internal particles are counted twice each while the contacts of the boundary particles with external particles are counted only once. This yields
\beq
z = \frac{2C_{II} + 2C_{IS} + C_{SE}}{{\cal{N}}} \ .
\label{zFric2D}
\eeq
The forces at the $C_{SE}$ contacts comprise the external loading on $\Gamma$ and are regarded as known boundary loading for the purpose of determining the intra-$\Gamma$ forces. These boundary forces are also presumed to be balanced (otherwise the assembly would not be static). The contacts $C_{II}$ and $C_{IS}$ transmit two force components each, giving $2\left(C_{II} + C_{IS}\right)$ unknowns to resolve within $\Gamma$. These are to be compared to the $3{\cal{N}}$ balance conditions. Defining $p_S\equiv{\cal{N}}_S/{\cal{N}}$, we then have
\begin{align}
J_A = &\frac{1}{{\cal{N}}}\left[2\left(C_{II} + C_{IS}\right) - 3{\cal{N}}\right]  \notag \\
= & z - 3 - \frac{C_{SE}}{{\cal{N}}}{{\cal{N}}} \notag \\
= & z - 3 - \alpha p_S \ ,
\label{Eqs_2D_f}
\end{align}
corresponding to the critical point shifting to \\
\beq
z_{c,A} = 3 + \alpha p_S \ .
\label{zcA}
\eeq

\noindent{\bf Frictionless non-discs in $d=2$}:\\
Using the same definitions as above, the number of equations is the same, $3N$, but only the force magnitudes at the internal contacts are unknown, $C_{II} + C_{IS}$. Then,
\begin{align}
J_B = &\frac{1}{{\cal{N}}}\left[\left(C_{II} + C_{IS}\right) - \left(3{\cal{N}}\right)\right]  \notag \\
= & \frac{z}{2} - \frac{C_{SE}}{2{\cal{N}}} - 3 \notag \\
= & \frac{z}{2} - 3 - \frac{\alpha}{2}p_S \ .
\label{Eqs_2D_nf}
\end{align}
The critical point in this case is at \\
\beq
z_{c,B} = 6 + \alpha p_S  \ .
\label{zcB}
\eeq

\noindent{\bf Frictionless discs in $d=2$}: \\
The number of unknowns is the same as in case B, but all the torque balance conditions are redundant, leaving only $2{\cal{N}}$ available equations. Therefore, 
\begin{align}
J_C = &\frac{1}{{\cal{N}}}\left[\left(C_{II} + C_{IS}\right) - 2{\cal{N}}\right]  \notag \\
= & \frac{z}{2} - \frac{C_{SE}}{2{\cal{N}}} - 2 \notag \\
= & \frac{z}{2} - 2 - \frac{\alpha}{2}p_S  \ .
\label{Eqs_2D_nfc}
\end{align}
The critical point in this case is at
\beq
z_{c,C} = 4 + \alpha p_S  \ .
\label{zcC}
\eeq
Thus, in all three cases, the change to the infinite critical value is by adding $\alpha p_S$.\\

The stability parameter $J$ can be used to define a phase field parameter in mechanically stable granular assemblies, $\Psi\equiv1 - H(J)$, with $H$ the Heavyside step function. $\Psi$ is unity in the marginally stable phase and vanishes in the over-connected phase. It can be used to develop phase-field simulations, in which it would determine the stress equations to use and where phase boundaries are.
 It is straightforward to extend the calculations of $J$ to three and higher dimensions, using the same rationale.

%===================================

\section{Conclusion}

To conclude, this paper should be regarded as a step toward a continuum stress theory of general mechanically stable GM, which goes beyond marginally stable states and the yield surface. The proposition is that real systems should be regarded as comprising two-phases: one marginally stable and the other over-connected. The conditions for marginal stability in large assemblies in arbitrary dimensionality and the first-principles formulation of isostaticity theory, including the explicit solutions to the stress field equations in $d=2$ have been reviewed briefly. 
A thought experiment was described which supports strongly the feasibility of the two-phase picture. In particular, it showed that there is a continuous spectrum of system structures that extends from the marginally stable state, through a general granular assembly and a porous medium, to a continuum uniform solid. 
To highlight the issues involved in deriving stress fields in two-phase systems, the problem was solved for a simple case -- a stack of plates of alternating phase. This problem also highlighted the constraints on the boundary conditions.
The critical-point-like nature of the marginally stable state has been used to extend the theory near this state. Specifically, a variation of the effective medium approximation (EMA) has been formulated for this problem and analysed. 
Finally, a quantitative stability parameter has been defined, which helps with the difficult problem of identifying the different phases and their boundaries within a given granular assembly. This parameter can be used for developing phase-field approaches to the problem. \\

Several points are worth discussing. One is the effects of gradients of $\overline{\overline{M}}$ on the stress chains typical length in the EMA method. 
The criticality of the marginally stable state is because a small local displacement of a particle is likely to break a contact, which destabilises the local structure by definition. This leads to local rearrangement, which causes another contact to break and so on. The long range rearrangement due to a small local perturbation is the analogue of a diverging response length near traditional critical points. While it is tempting to relate the rearrangement response to the stress and, in particular, to the typical length of stress chains, this relation holds only in media with relatively uniform fabric tensors, $\overline{\overline{M}}$.
This is because, as mentioned in section {\bf V}, spatial gradients of $m_{ij}$ give rise to secondary chains that split from the main chains and siphon stress away from them. Consequently, the stress attenuates along the main chain. The rate of this attenuation depends on the gradients magnitude along the chain and once the stress drops below some observability threshold, chains effectively terminate even though the medium is still ideally marginally stable and the rearrangement response is still very long-range. This is another manifestation of the decoupling between the stress and the strain in marginally stable media. \\

Another consideration enters this picture: isostaticity is a continuum theory and the EMA method requires an elementary volume over which the structure tensor is coarse-grained. This has two effects. One is that the gradients are milder on the coarse-grained scale and the other, that stress chains cannot be thinner than the linear size of an elementary volume. Both these effects counteract the shortening of the response length and must also be taken into account in structurally inhomogeneous systems. An investigation into this issue must also be part of the further development of the general stress theory.  \\

Another subtle issue is the following. In the solution for the uniform stress, (\ref{UniformStress}), whose full derivation is in the supplemental material\cite{SM1}, the $\sig_{yy}$ component of the boundary stress was taken to satisfy the stress-structure relation imposed by the local structure tensor, 
$\overline{\overline{M}}:\overline{\overline{\sigma}}=0$, and it is therefore a local function of $\sig_{xx}$ and $\sig_{xy}$. This may seem strange because one expects to be able to choose all the components of the boundary stress at will. However, there is no inconsistency! It has been shown that structure and the stress self-organise cooperatively~\cite{MaBl14,MaBl17,Jietal22}, namely, one cannot change without a corresponding change to the other. Self-organisation is a fundamental phenomenon GM, at least if the settling follows quasi-static dynamics. Thus, choosing a different value of $\sig_{yy}$ at some point on the boundary should have the effect of restructuring the contact network near that point, and that restructuring perturbation would propagate into the system as far as the stress response length. Such a self-organisation has been discussed and quantified to some extent in the literature~\cite{MaBl14,MaBl17,Bletal15}. Yet, there is no theory to predict the resultant modified structure tensor due to an arbitrary stress perturbation. It is likely that the structure would be most strongly modified close to the source of perturbation and unaffected very far from it, which means that gradients must develop. Once the structure has rearranged and the new structure tensor is known, the derivation of the stress field in the GM follows the same procedure that led to eq. (\ref{UniformStress}), albeit with coupling between the characteristics. 
Moreover, it is the inability to predict the structural response in marginally stable media to stress perturbations, which necessitated the tailoring of the boundary conditions to describe stress transmission from a marginally stable to elastic medium.  \\

While the discussion in this paper focused on two phases in static GM, it is interesting to note that two phases have also been discussed in the context of dense granular flows: plug regions, which are clusters of particles moving rigidly together, and plug-free regions, in which the velocity gradients are finite~\cite{daVinci1,daVinci2,KhRo17}. It is possible that, upon settling, the plug regions have a higher tendency to become the over-connected regions. This conjecture can be tested by measuring the correlation between a pre-settling particle belonging to a plug and its post-settling belonging to an over-connected particle. \\ 

Finally, there remain several hurdles in implementing this theory in practical modelling of natural systems and engineering applications. These include, but are probably not limited to, effective modelling of the constitutive fabric tensor $\overline{\overline{M}}$ on relevant lengthcales and determining the relative concentrations of the two phases. More work is needed to address these issues. However, the reward of such work cannot be overemphasised because a first-principles theory of real GM outside the yield surface has the potential to improve significantly predictability of models in a range of engineering disciplines. \\

\ni {\bf Conflict of interests}: \\
The author declares that he has no competing interests to report. \\

\ni {\bf Funding}: \\
This research did not receive any specific funding. \\

\ni {\bf Data availability statement}: \\
Data sharing not applicable to this article as no datasets were generated or analysed during the current study.


\begin{thebibliography}{99}

\bibitem{Archimedes} Archimedes, The Sand Reckoner (ca. 3rd century BCE) in {\it The Works of Archimedes}, ed. Thomas L. Heath (Cambridge University Press, UK 2009).
\bibitem{Re1885} O. Reynolds, Phil. Mag. Series 5. 20 (127): 469 (1885).
\bibitem{Ba41} R.A. Bagnold, {\it The Physics of Blown Sand and Desert Dunes} (Methuen, London, 1941).
\bibitem{EdOa89a} S.F. Edwards, R.B. Oakeshott, Physica {\bf D 38}, 88 (1989).
\bibitem{EdOa89b} S. F. Edwards, R. B. Oakeshott, Physica {\bf A 157}, 1080 (1989).
\bibitem{MeEd89} A. Mehta, S.F. Edwards, Physica, A {\bf 157}, 1091 (1989).
\bibitem{SeWu46} R.P. Seelig, J. Wulff, Trans. AIME {\bf 166}, 492, (1946).
\bibitem{Wa57} T. Wakabayashi, Proc. 7th Jpn. Nat. Cong. Appl. Mech., 153 (1957) (University of Tokyo Press 1958).
\bibitem{Da57} P. Dantu, Proc. 4th Int. Conf. Soil Mech. and Found. Eng., 144 (1957) (Butterworths 1957).
\bibitem{BaKi85} D.F. Bagster and R. Kirk, J. Powder Bulk Solids Technol. 1, 19 (1985).
\bibitem{HoBe97} D. Howell, R.P. Behringer, in {\it Powders and Grains 97}, Eds R.P. Behringer and J.T. Jenkins (Balkema, Rotterdam, 1997) pp 337.
\bibitem{OdKa98} M. Oda, H. Kazama, G\'eotechnique, {\bf 48}, 465 (1998).
\bibitem{Vaetal99} L. Vanel, D. Howell, D. Clark, R.P. Behringer and E. Clement, Phys. Rev. {\bf E 60}, R5040 (1999).
\bibitem{Maetal07} T. S. Majmudar, M. Sperl, S. Luding and R. P. Behringer, Phys. Rev. Lett. {\bf 98}, 058001 (2007).
\bibitem{Bl04} R. Blumenfeld, Phys. Rev. Lett. {\bf 93}, 108301-108304 (2004).
\bibitem{SM1} See Supplemental Material for the general solution of the isostatocity equations in $d=2$ and the solution of the isostaticity equations for the uniform boundary loading.
\bibitem{Mu63} N. I. Muskhelishvili, {\it Some Basic Problems of the Mathematical Theory of Elasticity} (Noordhoff, Groningen, 1963).
\bibitem{Wietal96} J. P. Wittmer, P. Claudin, M. E. Cates, J.-P. Bouchaud, Nature (London) {\bf 382}, 336 (1996).
\bibitem{Wietal97} J.P. Wittmer, M. E. Cates, P. Claudin, J. Phys. I (France) {\bf 7}, 39 (1997).
\bibitem{Caetal98a} M.E. Cates, J.P. Wittmer, J.-P. Bouchaud, P. Claudin, Phys. Rev. Lett. {\bf 81}, 1841 (1998).
\bibitem{Caetal98b} M.E. Cates, J.P. Wittmer, J.-P. Bouchaud, P. Claudin, Phil. Trans. Roy. Soc. A. {\bf 356}, 2535 (1998).
\bibitem{BaBl02} R. C. Ball, R. Blumenfeld, Phys. Rev. Lett. {\bf 88}, 115505 (2002).
\bibitem{Geetal08} M. Gerritsen, G. Kreiss, R. Blumenfeld, Phys. Rev. Lett. {\bf 101}, 098001 (2008).
\bibitem{BlMa17} R. Blumenfeld, J. Ma, Granular Matter19:29 (2017).
\bibitem{JaCo79} See, e.g., J.C. Jaeger, N.G.W. Cook, {\it Fundamentals of Rock Mechanics}, 3rd edn., (Chapman \& Hall, London 1979).
\bibitem{Tr64} H. Tresca, Comptes Redus Hebd. Seances l'Academies Des Sci. {\bf 59},  754 (1864).
\bibitem{vM13} R. v. Mises, Nachr. Ges. Wiss. Goett., Math. Kl. {\bf 1}, 582 (1913).
\bibitem{DrPr52} D.C. Drucker, W. Prager, Quarterly of Applied Mathematics {\bf 10}, 157 (1952).
\bibitem{NonRigid} R. Blumenfeld, in IMA Volume in Mathematics and its Applications, Vol. 141, pp 235-246: Modeling of Soft Matter, eds. Maria-Carme T. Calderer and E. M. Terentjev, (Springer-Verlag 2005).
\bibitem{Reetal13} J. Ren, J.A. Dijksman, R.P. Behringer, Phys. Rev. Lett. {\bf 110}, 018302 (2013).
\bibitem{Duetal14} G. D\"uring, E. Lerner, M. Wyart, Phys. Rev. E {\bf 89}, 022305 (2014).
\bibitem{Deetal14} E. DeGiuli, A. Laversanne-Finot, G. D\"uring, E. Lerner, and M. Wyart, Soft Matter 10, 5628 (2014).
\bibitem{MaBe05} T.S. Majmudar, R.P. Behringer, Nature {\bf 435}, 1079 (2005).
\bibitem{Zhetal08} J. Zhang, T.S. Majmudar, R.P.Behringer, Chaos 18:041107 (2008).
\bibitem{Zhetal10} J. Zhang, T.S. Majmudar, A. Tordesillas, R.P. Behringer, Granular Matter {\bf 12}, 159 (2010).
\bibitem{Daetal17} K.E. Daniels, J.E. Kollmer, J.G. Puckett, Rev. Sci. Instrum. {\bf 88}, 051808 (2017).
\bibitem{Zhetal19} Y. Zhao, H. Zheng, D. Wang, M. Wang, R.P. Behringer, New J. Phys. {\bf 21} 023009 (2019).
\bibitem{MoWa02} R.J. Mora, A.M. Waas, Proc. R. Soc. London A {\bf 458}, 1695 (2002).
\bibitem{Laetal17} H. Laubie, F. Radjai, R. Pellenq, F.-J. Ulm, Phys. Rev. Lett. {\bf 119}, 075501 (2017).
\bibitem{Raetal98} F. Radjai, D.E. Wolf, M. Jean, J.-J. Moreau, Phys. Rev. Lett., {\bf 80}, 61 (1998).
\bibitem{Muskh} N.I. Muskhelishvili, {\it Some Basic Problems of the Mathematical Theory of Elasticity} (Noordhoff, Groningen, 1963)
\bibitem{MaLi21} O. Mattei, M. Lim, J. Elasticity {\bf 144}, 81 (2021).
\bibitem{Cletal98b} P. Claudin, J.-P. Bouchaud, M.E. Cates, J.P. Wittmer, Phys. Rev. E {\bf 57}, 4441 (1998).
\bibitem{HoBe99} D. Howell, R.P. Behringer, C. Veje, Phys. Rev. Lett. {\bf 82}, 5241 (1999).
\bibitem{BeCh19} Robert P Behringer and Bulbul Chakraborty, Rep. Prog. Phys. {\bf 82} 012601 (2019).
\bibitem{Waetal19} M. Wang, D. Wang, J.E.S. Socolar, H. Zheng. R.P. Behringer,  Granular Matter {\bf 21}, 102 (2019).
\bibitem{MaBl14} T. Matsushima, R. Blumenfeld, Phys. Rev. Lett.{\bf 112} , 098003 (2014).
\bibitem{MaBl17} T. Matsushima, R. Blumenfeld, Phys. Rev. E {\bf 95} , 032905 (2017).
\bibitem{Jietal22} X. Jiang, R. Blumenfeld, T. Matsushima, "Coordinated Stress-Structure Self-Organization in Granular Packing", https://arxiv.org/abs/2208.06582.
\bibitem{Bletal15} R. Blumenfeld, S.F. Edwards, S.M. Walley, in {\it The Oxford Handbook of Soft Condensed Matter}, Eds. E.M. Terentjev and D.A. Weitz, (Oxford University Press, Oxford, UK, 2015), pp 167, ISBN-13: 978-0-19-966792-5.
\bibitem{daVinci1} R. Blumenfeld, S. F. Edwards, M. Schwartz, European J. Phys. E {\bf 32}, 333 (2010).
\bibitem{daVinci2} M. Schwartz, R. Blumenfeld, Granular Matter 13 , 241-245 (2011).
\bibitem{KhRo17}  P. Kharel, P. Rognon, Phys. Rev. Lett. {\bf 119}, 178001 (2017).
\bibitem{Bl20} R. Blumenfeld, Granular Matter {\bf 22}, 38 (2020).
\bibitem{Geetal08A} M. Gerritsen, G. Kreiss, R. Blumenfeld, Physica A {\bf 387}, 6263 (2008).

\end{thebibliography}
\end{document}